%Paper: hep-th/9301015
%From: A A Tseytlin <A.A.Tseytlin@damtp.cambridge.ac.uk>
%Date: Wed, 6 Jan 93 10:36 GMT
%Date (revised): Thu, 25 Feb 93 22:07 GMT

%%%%%%%%%%%%%%%%%%%%%%%%%%%%%%%%%%%%%%%%%%%%%%%%%%%%%%%%
\input harvmac.tex

\def \c { {{\rm cosh}  \ r}}\def \tt {{\tilde \t}}
\def \G {\Gamma}\def \tr { {\tilde \r}} \def \tk {{\tilde \k}}\def \k1 {{1\over
k}} \def \bh { \bar h } \def \ov { \over }
  \def \B { { \bar B }}
\def \Gl {\G_{loc}}
\def \O {\Omega }
\def \la { \tau }
\def \ra {\rightarrow}
\def \W { {\bar \o }}
 \def \in {\int d^2 z  }
\def \half{{\textstyle{1\over 2}}}

\def \ty { {\tilde y}}

\def \tx { {\tilde x}}
\def \a {\alpha}
\def \b {\beta}
\def \hi {\chi}
\def \sh {{\rm sinh \ }}
\def \Tr {{\ \rm Tr \ }}

\def \ln {{\rm \ ln \  }}
\def \det {{\ \rm det \ }}
\def \ch {{\rm cosh \ }}
\def \th {{\rm tanh  }}
\def \l {\lambda}
\def \1p {{1\over  \pi }}
\def \2p {{{1\over  2\pi }}}
\def \4p {{ {1\over 4 \pi }}}
\def \8p {{{1\over 8 \pi }}}
\def \P^* { P^{\dag } }
\def \p {\phi}

\def \m {\mu }
\def \n {\nu}
\def \ep {\epsilon}
\def\g {\gamma}
\def \r {\rho}
\def \k {\kappa }
\def \d {\delta}
\def \o {\omega}
\def \s {\sigma}
\def \t {\theta}

\def \fourth {{\textstyle{1\over 4}}}

\def \e#1 {{\rm e}^{#1}}
\def \const {{\rm const }}

\def \eq#1 {\eqno {(#1)}}
%%%%%%%%%%%%%%%%%%%%%%%%%%%%%%%%%%%%%%%%%%%%

\def \bd  { \bar \del }
\def \A  { \bar A }

%%%%%%%%%%%%%%%%%%%%%%%%%%%%%%%%%%%%%%%

\def \o {\omega}

\def \p {\phi}
\def \ep {\epsilon}
\def \s {\sigma}

\def \r {\rho}
\def \d {\delta}
\def \l {\lambda}
\def \m {\mu}
\def \g {\gamma}
\def \n {\nu}

\def \fourth {{1\over 4}}

\def \e#1 {{\rm e}^{#1}}
\def \const {{\rm const }}

\def \B {{\bar B}}

\def \half { { 1\over 2 }}
\def  \ee {{\rm e}^}
\def \J {\bar J }
\def\np {  Nucl. Phys. }
\def \pl { Phys. Lett. }
\def \mpl { Mod. Phys. Lett. }
\def \prl { Phys. Rev. Lett. }
\def \pr  { Phys. Rev. }

\def \cmp { Commun. Math. Phys. }
\def \ijmp { Int. J. Mod. Phys. }

\Title{\vbox{\baselineskip14pt{\hbox{Imperial/TP/92-93/10} }
{\hbox{hep-th/9301015}}}}
{\vbox{\centerline{  Effective action of  gauged WZW model }
\vskip2pt\centerline{  and exact string  solutions}}}

\centerline{ A.A. Tseytlin
\footnote{$^*$}
{e-mail: tseytlin@surya3.cern.ch and tseytlin@ic.ac.uk }
\footnote{$^\star$} {Present address: TH, CERN, Geneve 23,
 CH-1211, Switzerland. On leave of absence from the Department of
Theoretical Physics, P. N. Lebedev Physics Institute, Moscow 117924, Russia.}}
\centerline{\it  Theoretical Physics Group }
\centerline {\it  Blackett Laboratory}
\centerline{\it Imperial College}
\centerline{\it  London SW7 2BZ, United Kingdom }

\baselineskip=20pt plus 2pt minus 2pt
\vskip .3in

%abstract
\noindent
We suggest how to derive the exact (all order in $\a'$) expressions for
 the background fields  for string solutions  corresponding  to  gauged WZW
models directly at the $2d$ field theory level. One is first to   replace the
classical gauged WZW action by the quantum effective one and then to integrate
out the gauge field.  We find the explicit expression  for the gauge invariant
non-local effective action of the gauged WZW model.  The two terms
(corresponding to the group and subgroup) which appear with the same
coefficients in the classical action get different $k$-dependent
coefficients in the effective one.  The procedure of integrating out the gauge
field is considered in detail for the $SL(2,R)/U(1)$ model and the exact
expressions for the $D=2$ metric and the dilaton (originally  found   in the
conformal field theory approach) are reproduced.

 \bigskip
\it{ to appear in  Nuclear Physics B    }
\Date{12/92} %replace this line by \draft  for preliminary versions
             %or specify \draftmode at some point

\baselineskip 20pt plus 2pt minus 2pt
%if you want double-space, use e.g. \baselineskip=20pt plus 2pt minus 2pt

 %\draftmode \baselineskip 12pt plus 2pt minus 1pt
\noblackbox
%%%%%%%%%%%%%%%%%%%%%%%%%%%%%%%%%%%%%%%%%%%%%%%%%%%%%%%%%%%%%%%%%%%%%%%
\lref \gwz { P. Di Vecchia and P. Rossi, \pl B140(1984)344;     K. Bardakci, E.
Rabinovici and
B. S\"aring, \np B299(1988)157;
 K. Gawedzki and A. Kupiainen, \pl B215(1988)119;
\np B320(1989)625. }

\lref \sen {A. Sen, preprint TIFR-TH-92-57. }

\lref \bcr {K. Bardakci, M. Crescimanno and E. Rabinovoici, \np
B344(1990)344. }
\lref \Jack {I. Jack, D.R.T.  Jones and J. Panvel, Liverpool preprint
LTH-277 (1992). }
\lref \zam  { Al. B. Zamolodchikov, preprint ITEP 87-89. }

\lref \hor {J. Horne and G. Horowitz, \np B368(1992)444. }
\lref \tse { A.A. Tseytlin, \pl B264(1991)311. }

\lref \ishi { N. Ishibashi, M.  Li and A. Steif, \prl 67(1991)3336. }
\lref  \kumar  { M. Ro\v cek and E. Verlinde, \np B373(1992)630; A. Kumar,
preprint CERN-TH.6530/92;
 S. Hussan and A. Sen,  preprint  TIFR-TH-92-61;  D. Gershon,
preprint TAUP-2005-92; X. de la Ossa and F. Quevedo, preprint NEIP92-004; E.
Kiritsis, preprint LPTENS-92-29. }

\lref \rocver { A. Giveon and M. Ro\v cek, \np B380(1992)128. }
\lref \frts {E.S. Fradkin and A.A. Tseytlin, \np B261(1985)1. }
\lref \mplt {A.A. Tseytlin, \mpl A6(1991)1721. }
\lref\bn {I. Bars and D. Nemeschansky, \np B348(1991)89.}
\lref \shif { M.A. Shifman, \np B352(1991)87.}
\lref\wittt { E. Witten, \cmp 121(1989)351; G. Moore and N. Seiberg, \pl
B220(1989)422.} \lref \chernsim { E. Guadagnini, M. Martellini and M.
Mintchev, \np B330(1990)575;
L. Alvarez-Gaume, J. Labastida and A. Ramallo, \np B354(1990)103;
G. Giavarini, C.P. Martin and F. Ruiz Ruiz, \np B381(1992)222; preprint
LPTHE-92-42.}
\lref \shifley { H. Leutwyler and M.A. Shifman, \ijmp A7(1992)795. }
\lref \polwig { A.M. Polyakov and P.B. Wiegman, \pl B131(1984)121.  }
\lref \polles { A. Polyakov, in: Fields, Strings and Critical Phenomena,
Proc. of  Les Houches 1988, eds E. Brezin and J. Zinn-Justin
 (North-Holland, 1990).   }
\lref \kutas { D. Kutasov, \pl B233(1989)369.}
\lref \karabali { D. Karabali, Q-Han Park, H.J. Schnitzer and Z. Yang, \pl
B216(1989)307;  D. Karabali and H.J. Schnitzer, \np B329(1990)649. }
\lref \ginq {P. Ginsparg and F. Quevedo,  \np B385(1992)527. }
\lref \gko  {K. Bardakci and M.B. Halpern, \pr D3(1971)2493;      P. Goddard,
A. Kent and D. Olive, \pl B152(1985)88.  }
\lref \dvv  { R. Dijkgraaf, H. Verlinde and E. Verlinde, \np B371(1992)269. }
\lref \kniz {  V. Knizhnik and A. Zamolodchikov, \np B247(1984)83. }

\lref \witt { E. Witten, \cmp 92(1984)455.}
\lref \wit { E. Witten, \pr D44(1991)314.}
\lref \anton { I. Antoniadis, C. Bachas, J. Ellis and D.V. Nanopoulos, \pl B211
(1988)393.}
\lref \bsfet {I. Bars and  K.Sfetsos, \pr D46(1992)4510; preprint
USC-92/HEP-B3, \pl (1992).}

\lref \ts  {A.A.Tseytlin, \pl B268(1991)175. }
\lref \kir {{E. Kiritsis, \mpl A6(1991)2871.} }

\lref \shts {A.S. Schwarz and A.A. Tseytlin,  preprint Imperial/TP/92-93/01,
Nucl.Phys. To appear. }
\lref \bush { T.H. Buscher, \pl B201(1988)466.  }

\lref \plwave { D. Amati and C. Klim\v cik, \pl B219(1989)443; G. Horowitz and
A. Steif, \prl 64(1990)260.}
\lref\bsft { I. Bars and K. Sfetsos, \pl B277(1992)269; \mpl A7(1992)1091;
 \pr D46(1992)4495;
I. Bars, preprint USC-92/HEP-B5. }
\lref \div  { P. Di Vecchia,  V. Knizhnik, J. Peterson and P. Rossi, \np
B253(1985)701;
 A. Redlich and H. Schnitzer, \pl B167(1986)315; B193(1987)536(E);
E. Bergshoeff, S. Randjbar - Daemi, A. Salam, H. Sarmadi and E. Sezgin,
\np B269(1986)77; J. Fuchs, \np B318(1989)631;
 H. Schnitzer, \np B324(1989)412.  }
\lref\bs { I. Bars and K. Sfetsos, Univ. Southern California preprint
USC-93/HEP-B1.   }
\lref\bst { I. Bars,  K. Sfetsos and A.A. Tseytlin, unpublished. }
\lref\bb  { I. Bars, \pl B293(1992)315. }
\lref \ttt {A.A. Tseytlin, preprint CERN-TH.6804/93.}
 %%%%%%%%%%%%%%%%%%%%%%%%%%%%%%%%%%%%%%%%%%%%%%%%%%%%
\newsec {Introduction }
Solutions of (tree level) string field equations   are usually represented in
terms of conformal invariant $2d$ theories.  In the context of particle
theory applications of string theory one is interested in string vacuum
backgrounds which  are direct products of  non-trivial  euclidean ``internal"
part  and flat $4d$ Minkowski space-time.
At the same time, interpreting string theory as a theory of (quantum) gravity
one needs to study  solutions which have  curved  space-time part, i.e.
which should be described by  conformal theories with Minkowski signature
of a target space.  In the context of gravitational applications of string
theory it is important to  be able to go beyond the leading orders of expansion
in $\a'$ (the leading order solutions may not   correctly describe
the behaviour in the strong curvature regions near
singularities, etc). Since the string effective  equations  contain terms of
arbitrarily high order in $\a'$ their  solutions will in general be non-trivial
functions of $\a'$.

Given that   the exact  form of the string effective action is not known
it is difficult to expect that  such solutions can be found directly in the
perturbative sigma model framework (there are of course special
solutions for which  higher order corrections cancel automatically  as in the
cases of  group spaces  \witt \ or   `plane wave'  backgrounds \plwave).

A  number of new solutions of string equations  can in principle be
constructed by  starting with gauged WZW models \bn\wit\ and  also by  using
duality transformations of the known solutions (see e.g.\sen\ for a review).
 However, both the standard procedure of extracting
space-time backgrounds by integrating out gauge fields \wit
\foot {The gauged WZW model is conformal
only under a special prescription in which the group field $g$ and the gauge
field $A$ are treated on an equal footing. Integrating out $A$ while keeping
$g$ classical gives  a model which is conformal only in the one-loop
approximation. Though the formal transformations of the path integral may look
preserving equivalence, to  justify them one needs to introduce
sources/backgrounds (as one is  to do in order to compute e.g. $\b$-functions)
and also to account for  the effects of Jacobians resulting from
integrations.} and (the leading order form of) duality transformations
preserve conformal invariance  only in the  leading $\a'$  approximation.
\foot {The question of higher order  corrections to sigma model duality
transformations was addressed in \mplt. Given that the exact form of these
transformations is not known it is important  to develop other methods of
establishing the exact form of solutions.    Since the leading order
backgrounds corresponding to gauged WZW models (with abelian subgroup being
gauged) can be represented as duality rotations
 of  ungauged WZW  models \kumar\ understanding  of how to find the  exact
 backgrounds may shed light on exact form of duality transformations.}

It should be possible to determine the exact form of the background  fields
in the case when one knows explicitly the conformal theory corresponding to a
particular solution. To give a space-time interpretation to a
conformal theory (i.e. to find the  couplings of the corresponding sigma
model) is, in general,  a non-trivial problem. One possible strategy is to
consider perturbations of a given conformal theory by marginal operators
considered as   ``probes" of   geometry, interpreting  the  linear
$L_0$-constraints as ``Klein-Gordon" equations  and trying to extract the
values of the  background fields from the
 explicit form of the differential operator $L_0$ (see e.g. \anton). This
approach  first applied to derivation of exact background for a particular
($SL(2,R)/U(1)$) model in \dvv\  was  developed in full generality
 (and   with important understanding  of a possibility to  determine global
 geometry) in  \bsfet.
   Though   being consistent with perturbative solution of sigma model
conformal invariance conditions (as was checked in the  $SL(2,R)/U(1)$ case
\dvv\  up to three \ts\ and four \Jack\ loops) the ``$L_0$ - approach" is,
however,  rather heuristic being based on a number of implicit assumptions
about the structure of the string effective action (or about  how $L_0$ should
depend on the dilaton, how the antisymmetric tensor should enter
 equations for higher level modes, etc).

Below we shall suggest   the  explicit method  of  determination  of the
background geometry  directly at the  $2d$ field theory level.
 The main idea  is   first to replace the classical gauged WZW action by
a quantum effective one (with all $\a'$ or $1/k$ corrections already accounted
for) and $then$ to eliminate the gauge field as in \wit. The resulting sigma
model should  have the $exact$ space-time fields as its couplings.

The reason why  one   should be able
to find the exact background in the gauged WZW model directly in the $2d$ field
theory framework is that  being soluble in the conformal operator approach the
model should be exactly solvable  also  in the Lagrangian approach, i.e.
its effective action  should be explicitly computable as a
{\it non-trivial} function of the  parameter $k$. As we shall see in
Sec.2  the effective action in  the ungauged WZW theory
is just equal to the  classical one with $k$ replaced by $k + \ha c_G$
($c_G$ is the value of the second Casimir
operator in the adjoint representation of $G$)
so that the
corresponding exact  metric is just equal to the  classical group space
metric   multiplied by  $k + \ha c_G$)\foot {The presence of this  extra
factor in the metric is consistent with the ``hamiltonian"  argument based on
interpreting the $L_0$ - operator as  a Klein-Gordon operator in a
background. }. At the same time,  the classical action  of the gauged WZW
model  (Sec.3) contains two  different structures the coefficients of which
get  ``renormalised" in  different ways \bst. As  a  result,   the effective
action  and hence the corresponding background  fields  are  non-trivial
functions of $k$ (or $\a'$).\foot {The backgrounds found  in \dvv\bsfet\
provide first known examples of string solutions
 that are non-trivial
functions of $\a'$.  The $\a'$ -  corrections, though absent in the
supersymmetric case \Jack\bsfet , are still present  in the heterotic one
\bsfet\bb . }

In Sec.4 we shall consider the $SL(2,R)/U(1)$ gauged WZW model and show
that eliminating   the gauge field from the corresponding effective  action
one reproduces the exact expressions for the  background  metric and dilaton
found  indirectly  within  the conformal field theory ($L_0$-operator) approach
in \dvv. It is interesting to note  that if one modifies the coefficient of the
$A^2$ term in the  classical action of the $SL(2,R)/U(1)$ model
$A^2$ $\ra (1+ { 4\over k-2 })A^2 $ and then eliminates $A$ from the action
 one  finds the exact metric of \dvv. However, this naive ansatz
for the ``quantum" action
 is not gauge invariant and
does not  reproduce the  exact  form of the dilaton background \dvv\ts.
The  correct effective action   found according  to the general procedure
described in  Sec.3   corresponds to  modifying (by ${ 4\over k-2 }$) of
only the coefficient of the square of  the $transverse$ part of $A$.

Sec.5 contains some concluding remarks.
In Appendix we discuss  some  actions which  are related to
effective action of the $SL(2,R)/U(1)$ model by
duality transformations.

 %%%%%%%%%%%%%%%%%%%%%%%%%%%%%%%%%%%%%%%%%%%%%%%%%%%%
%%%%%%%%%%%%%%%%%%%%%%%%%%%%%%%%%%%%%%%%%%%%%%%%%%%%%%%%%%%%%%%%%%%%

\newsec{ Effective action in  WZW model}
Our aim will be to  show that the   effective action of WZW
theory \witt\polwig\  is equal to the classical action with the level $k$
replaced by $k + \half c_G$. Though such shift  of $k$ is
a well-known quantum effect in the current algebra  relations
{\kniz }  its origin within   the
Lagrangian  field theory  approach  does not seem to be widely understood.
Analogous shift of $k$
in  the effective action of related $3d$ Chern-Simons theory \wittt\  was
discused, e.g., in \chernsim\shif.  According to \shif\ the shift of $k$
in CS theory is essentially a one-loop phenomenon which takes place when one
passes from the operator Wilson effective action to the standard $c$-number
one.

 An interesting  analysis of  WZW theory   from the  perturbative
field theory  point of view   was recently  presented in \shifley.
It was shown that the generating functional $W(B)$ for correlators of chiral
currents is given by its classical expression, i.e. does not receive quantum
corrections. The correlator of two currents for zero external field is equal
to the classical expression  proportional to $k$. For non-zero external field
it contains both connected ($ \d^2 W \over \d B(z) \d B(z')$) and disconnected
($ -{\d W \over \d B(z)} {\d W \over \d B(z')}$) parts. Though $W$ is
classical,  in the limit $z\ra z'$ the connected part (proportional to the
Planck constant)  takes  effectively the form of a one-loop  contribution.
One term in the latter  renormalizes the  disconnected contribution (which is
proportional to the energy-momentum tensor). This is the origin  of the
shift  $k \ra k + \half c_G$
 in the  corresponding OPE $\  J(z)J(z') \sim   ( k + \half c_G) T(z) $.

The results of \shifley\  suggest  that the shift of $k$  which from one
point of view is due to a  subtle  relation  between  some connected and
disconnected graphs  may  be
reinterpreted as originating from a {\it one-loop   determinant}.
In fact, as we shall argue  below
 the shift of $k$ in WZW theory can be  understood as
appearing from  a     determinant \polwig\   identified
as being  a Jacobian of a change of variables which is necessary to perform in
order to define the quantum effective action by a {\it Legendre }
transformation.

 Consider the following path integral\foot { Though we  shall use complex
euclidean  coordinate notation ( $ \del_a \del^a = 4 \del \bd $ , etc)   we
understand that in order to satisfy the natural reality conditions (especially
in the context of gauged WZW theory) one should  actually consider the case of
Minkowski  signature \karabali. }
 $$
  \ee{-W(B) } =  \int [dg] \  \e{ - S(g)  +  B\bar J (g) } \ \ , \ \ \ \   \
 S=
{k }  I(g) \ \ ,\eq{2.1} $$ $$  \ \  I \equiv  {1\over 2\pi }
\int d^2 z  \Tr (\del g^{-1}
\bd g )  +  {i\over  12 \pi   } \int d^3 z \Tr ( g^{-1} dg)^3   \ \ , \
\eq{2.2}
$$
 $$ B\J (g) =  {k\over \pi }  \int d^2z \Tr (B \bd g g^{-1} ) \ \ . \eq{2.3} $$
Using the identity \polwig
$$ I(ab) = I(a) + I(b) - {1\over \pi }  \int d^2z \Tr (a^{-1} \del a \ \bd b
b^{-1} ) \ \ ,  \eq{2.4} $$
it  is easy to see that the generating functional for the  correlators of the
current $\J$ is given by \polwig\polles
 $$   W(B) = - {k  }  I(u) \ \ , \
\  \ \ B = u^{-1} \del u  \ \ . \eq{2.5} $$
Equivalently, that means that  $W$ does not receive quantum
corrections, i.e. it is  equal to  the classical action  evaluated on the
classical solution depending on $B$ \shifley.

Given that   $B$ plays the role of an external
current associated with the composite field $\J$ (i.e.  $u$ is a ``source"
corresponding to $g$)   to get the effective action  for $g$ itself we still
need
to make the  Legendre transformation  of $W(B)$.  Instead of just solving the
classical equations for $B$  let us   consider  the  path integral
over it  and define the functional $\G (g)$     $$
  \ee{- \G (g) } =  \int [dB] \  \e{ - W(B)  +  B\bar J (g) } \ \ .  \
\eq{2.6} $$
Since  the quantum corrections in the corresponding path integral
effectively cancel out \shifley\  the functional  $\G$ will represent the
required Legendre transform of $W$.  As follows from (2.1) $$\ee{- \G (g) } =
\int [dg'] \  \e{ - S(g') }  \ \delta [  \J (g') - \J  (g) ]
 \ \ ,  \  \eq{2.7} $$
 so that $\G$ has a natural  ``effective action" interpretation. To compute
(2.6)  we change the variable  $ B = u^{-1} \del u  $ to $u$. The
resulting Jacobian is a determinant of a chiral operator. Since the original
model is non-chiral the determinant  should be defined in a  left-right
symmetric  way  (i.e. in a way
preserving vector gauge symmetry) as the square root of the
product \polwig\polles\foot {
  The chiral determinant is not
unambiguously defined. It was also suggested \kutas\ that in the chiral case an
additional wave function renormalisation should accompany the change $B \ra u
\ $
 $$ \int [dB] \  \exp{[ - W(B)  +  B\bar J (g) ]}  =
 N \int [du] \  \exp {[ (k+c_G) I(u)  +  p  B(u)\bar J (g) ]}  \ ,
 $$
where  $p = \sqrt {(k+ c_G)/k } $.}
 $$
 \det ( \del + [B, \ ] )  \det ( \bd + [\B, \ ] )  =
 \exp { [   c_G I(v^{-1}u) ] } \det \del \det \bd \ \ ,
\eq{2.8} $$
$$  \ B= u^{-1} \del u \
\ , \ \ \ \
 \  \B = v^{-1} \bd v   \ \ .
$$
In the present case of  $v=1$ we get
 $$ [d B ]  = [du] \det ( \del + [B,\  ] )  = [du] \
\exp { [ \ha c_G I (u) ] } \ (\det \del\bd )^{1/2}
 \ \  .  \ \eq{2.9} $$
Then
 $$ \ee{- \G (g) } =  N \int [du] \  \exp { [ (k + \half c_G) I(u)   +  B(u)
\bar J (g) ]} \ \ .  \  \eq{2.10} $$
As a result,  we finish with (cf.(2.1),(2.5);
 $\  k \ra  - (k+
\half c_G) $)\foot { We  suppress  the wave-function renormalisation
corresponding to $ k\bd g g^{-1}\ra (k + \ha c_G )\bd g g^{-1} . $
Similar effective actions were discussed in the context of
$sl(2,R)$ approach to $2d$ quantum gravity in\zam\ . }
$$  \G (g) = (k + \half c_G) I(g)  \ \ . \eq {2.11} $$
The shift of $k$ in the effective action is perfectly consistent  with the
presence of the shifted $k$ in the quantum currents or in the quantum equations
of motion in \kniz\ ($ \ (k + \half c_G)\bd g =\pi  {:{\bar J }g:}\ $,
etc).
  Given that
the energy-momentum tensor $T$  should be represented  by the  variation of the
effective action over the background $2d$ metric  eq.(2.11) is  also consistent
with the standard  operator expression $ T(z) = {1 \over k + \half c_G }
:J^2:\  $.

The expression (2.11) for the effective action  implies that the exact target
space metric of the corresponding conformal sigma model is given by the group
space metric multiplied by  $k + \half c_G$.  This result  is  the same as
 one finds     by interpreting  the $L_0$ operator of
the WZW conformal theory as a Klein-Gordon operator of a space-time theory
 \anton\dvv\bsfet.  Though a particular value of the overall  constant
factor in  the action (or in the target space fields)  does not seem
important,  the observation  that the effective action of the WZW model is
multiplied by the shifted $k$ will have  a  non-trivial implication for the
structure of the effective action in the gauged WZW model.

%%%%%%%%%%%%%%%%%%%%%%%%%%%
%%%%%%%%%%%%%%%%%%%%%%%%%%%
%%%%%%%%%%%%%%%%%%%%%%%%%%%%%%%%%%%%%%%%%%%%%%%%%%%%%%%%%%%%%%%%%%%%%%%%

\newsec { Effective action in gauged WZW model }
Let us start with a  review  of the path integral quantisation of gauged WZW
model following \karabali (see also \gwz).
The action
$$ \ \  I = {1\over 2\pi }
\int d^2 z  \Tr (\del g^{-1}
\bd g )  +  {i\over  12 \pi   } \int d^3 z \Tr ( g^{-1} dg)^3   \ \ , \
\eq{3.1}
$$
is invariant under the global $G_L\times G_R$  transformations.  The action
invariant under the $holomorphic$ vector gauge transformations $ g \ra
u^{-1}(z) g u(\bar z) $ (where $u$ belongs to a  vector subgroup $H$)  is
$$ I_0(g,A) = I(g) - {1\over \pi }
 \int d^2 z \Tr \bigl(  A\,\bd g g\inv
- \bar A \,g\inv\del g - g\inv A g \bar A
\bigr)\  \ .  \eq{3.2} $$  To  cancel the anomaly in the effective action one
is
to add the counterterm $\Tr(A \A)$ (see e.g. \ginq). The final  gauged
action\foot { In  the case when one gauges  the axial $U(1)$  subgroup  the
$A\A$ term should be added with the opposite sign, or, equivalently (after
changing $A\ra - \A$) the  minus signs in the brackets in the second term in
(3.3) should be replaced by the plus signs.
Let us note that there exists a more general left--right asymmetric way of
gauging found in the second reference in \bsft. }
 $$ I(g,A) = I(g)  - {1\over \pi }
 \int d^2 z \Tr \bigl(  A\,\bd g g\inv -
 \bar A \,g\inv\del g - g\inv A g \bar A  + A \A \bigr)\  \   \eq{3.3} $$
is invariant under  the  standard  vector  gauge transformations:
$$ g \ra u^{-1} g u \ , \  \ A \ra u\inv ( A + \del ) u \ , \ \
 \A \ra u\inv ( \A + \bd ) u \ , \ \  \ \ u = u (z, \bar z)  \ .$$
The invariance of
(3.2) under  the holomorphic gauge transformations   indicates that  the $ A\A$
term  plays a special   role  in (3.3), i.e.  that (3.3)  should  be
considered as a sum of $two$ different structures: $I_0$ (3.2)  and  $\1p \int
d^2z \Tr (A\A ) $.  In fact,  as we shall see
below,   the coefficients of $I_0$ and  $A\A$  will  become  different    in
the
quantum effective action corresponding to (3.3).

  Parametrising  $A$ and $\A$ in terms of $h$ and $\bh$  which
take values in  $H$
 and transform as $ h \ra u\inv h \ , \ \ \bh \ra   u\inv \bh $,
$$ A = h \del h\inv \ \ , \ \ \ \A = \bh
\bd \bh\inv  \ \ , \eq{3.4} $$
and using the identity  (2.4) we can represent (3.3)  as the  difference of the
two explicitly gauge invariant terms  (given by WZW actions  corresponding to
$G$
and $H$) \karabali
$$ I(g,A) = I (h\inv g \bh ) -  I (h\inv \bh)  \ \ . \eq{3.5} $$
Starting with  the path
integral (with the Haar measure for $g$ and the canonical measure for $A, \A$
including  gauge fixing)
$$ Z = \int [dg] [dA][d\A] \ \ee{ - k I(g,A) }      \ \  \eq{3.6} $$
and changing the variables using  (2.8) \polwig , i.e.
$$ \det (\del + [A, \ ] ) \det (\bd + [ \A , \ ] )
= \exp{ [ c_H I(h\inv \bh ) ]} \det \del \det \bd \ \ , \ \eq{3.7} $$
we get
$$Z = \int [dg] [dh][d\bh] \exp{ [ - k I (h\inv g \bh ) + (k + c_H)  I (h\inv
\bh)    ]    } \ \ .
\eq{3.8} $$
The last expression implies that $ I(g,A)$  can be quantised as the sum of
the two WZW theories for the groups $G$ and $H$ with levels $k_G=k$ and
$k_H=-(k+c_H)$.  According to our discussion  in Sec.2 to get the
 effective action  of WZW theory one is to replace $k$ by
$k + \half c_G$ (see (2.11)). This suggests that the effective action in the
gauged WZW theory  is explicitly computable and is given by
 $$ \G (g,A) = (k + \half c_G)I (h\inv g \bh ) - (k + \half  c_H)  I (h\inv
\bh)        \ \ .  \eq{3.9} $$
The contribution of the Jacobian (3.7) is
essential in order to make (3.9)  a natural generalisation of  both
the classical  gauged WZW action (3.5)  and the effective action of the WZW
theory  (2.11) (note that $-(k +c_H ) + \half c_H = - (k+ \half c_H)$).
The structure of (3.9) is in correspondence with that  of the
holomorphic energy-momentum   tensor  operator  in the conformal field theory
approach \gko
$$ T(z) = {1 \over k + \half c_G } : J_G^2 :\  - \  {1 \over k + \half c_H } :
J_H^2 : \ \ .$$
It should be noted that in (3.9)   we have
ignored  field renormalisations.\foot { The conjecture about the
structure (3.9) of the effective action  emerged as a result of collaboration
with I. Bars and K. Sfetsos \bst.} The latter, however, are not
relevant for our aim of deriving the  exact backgrounds corresponding to
the  $2d$ action  in which $h, \bar h$ are  already eliminated.\foot {
 The true reason why the effects of field renormalisations can be ignored
here is that they introduce extra non-local terms in the action which
can be ignored in the process of identifying the  couplings  in the clasiical
sigma model action, see Note Added. }

One can now re-write (3.9) in terms of the original variables $A, \A$
(note that since $\G$ is the effective action,  its arguments
should be treated as classical fields).
Using (3.5) we can
represent  the effective action (3.9) in the form $$\G(g,A) = (k+ \half c_G)
[\  I (g, A ) + { (c_G- c_H) \over 2(k+ \half c_G) } \O(A) \ ]
  \ \ , \eq{3.10} $$
where
$$ \O (A) \equiv I(h\inv \bh) = \o (A) + \W (\A) + \1p \int d^2 z \Tr (A\A)
   \ \  , \eq{3.11} $$ $$ \ \ \ \o (A) \equiv I(h\inv)\ , \ \ \ \ \W(\A)
\equiv  I(\bh)\ \
 . $$
$\O$  is a non-local {\it gauge invariant} functional of $A, \ \A $ \foot {As
it is clear from (3.7) $\o$  and $\W$ are  effective actions  corresponding to
chiral (fermion) determinants.}
which   represents the gauge-invariant part of $\int d^2 z \Tr (A\A)$
 (note that it is
 the coefficient of the gauge invariant part of the $A\A$ term  in the
classical action (3.3)  that got modification different from that of the
rest of the  coefficients  in the  action).

Let us look at the structure of the two terms in  (3.5) in more
detail.  Using the identity (2.4) and the definition (3.2) of $I_0$  we get
 $$ I(h\inv g \bh ) = I_0 (g,A)  + I(h\inv) + I(\bh) \ \ , \ \
\ \eq{3.12} $$ $$  I(h\inv \bh) = \1p \int d^2 z \Tr (A\A)  +   I(h\inv) +
I(\bh) \ \ . \eq{3.13} $$ While the non-local terms $I(h\inv) + I(\bh)$  cancel
out in the  classical action (3.5) they  survive in the effective one
(3.9),(3.10).

In the  case   when $H$ is abelian  the functional $\O$  (3.11) has the
following explicit form
 $$ \O = \2p \int  \Tr  F {1 \over \del \bd } F  \ \  \ , \ \ \ \ \ F= \bd A -
\del \A \ \ , \ \eq{3.14} $$
or, eqivalently,
$$ \O = \1p \in \Tr ( A\A  - \half A{\bd\over \del} A - \half \A { \del \over
\bd}
\A  )  \ \  .\eq{3.15} $$
Since the integrand of  $\O$ (3.14),(3.15) is   just  the square
of the transverse part of $A_a$   (i.e.  $\O$ is proportional to  the effective
action of the Schwinger model) it can be rewritten  also as  the local $A_a^2$
term  minus the  square of the gauge-dependent longitudinal part of $A_a$
 $$ \O
= \1p \in \Tr (  2 A\A  + \half  F'{1 \over \del \bd } F'  )\ \ , \ \ \  \
F'\equiv  \bd A +  \del \A \ \ . \ \eq{3.16} $$
Note that the coefficient of the  $ A\A$ term in (3.16) has doubled.
The $local $ part of the effective action (3.10) in the abelian
case ($c_H=0$)  then takes the form
$$\Gl(g,A) = (k+ \half c_G) \big[ \  I (g, A ) + \1p { { c_G \over  (k+ \half
c_G) }}  \in \Tr (   A\A  )\  \big]
 $$
$$ = (k+ \half c_G) \big[ \ I_0 (g,A) - \1p (1 -
{ c_G \over {k+ \half c_G } } )
\in \Tr (   A\A  )\  \big] \ \ . \eq{3.17} $$
In general, it is possible to choose a gauge in which  $F'=0$  so  that $\G$
in (3.10) is equal to (3.17).  That means that (normalised) correlators of
{\it gauge-invariant} operators computed using  the full $\G$ (3.10) and its
local part $  \G_{loc}$ will be the same.

Having found the effective action $\G (g, A )$  one  may  follow
 \wit ,  solving  for the gauge field and eliminating  it from the action (i.e.
integrating it  out in semiclassical approximation).  This can be done in the
two equivalent ways depending on whether one fixes a gauge on $g$ or on $A_a$.
If we start directly with (3.10) and eliminate $A_a$ we will get an effective
action $\G (x)$  which will still be gauge invariant, i.e. will depend on
gauge-invariant  combinations $x (g) $ of the fields   $g$.
Elimination of the   gauge field is  more  straightforward if one  first
fixes the gauge on $A_a$  in which  $\G$ (3.10) reduces to its local part
(3.17). Since $\Gl$ is local and quadratic in $A_a$ the result will be a local
action $\G' (g)$, which, however, will no longer  be gauge invariant, i.e.
it will depend  on all components of $g$.  Integrating out the gauge degrees of
freedom will lead us back to $\G (x)$.

The resulting model should  be  conformally invariant to all loop
orders.  A non-trivial question is whether its action $\G (x)$ will be local.
Since  the effective action (3.10) is non-local in the $A_a$-sector, it
remains non-local in the gauge $g=x$ so that it is not {\it a priori } clear
why integrating out $A_a$  will give us   a  local action.  If we start with
(3.17) we get a local action $\G' (g)$  which may, however, reduce to a
non-local once one decides to eliminate the gauge degrees of
freedom.\foot { One should, in fact, identify the effective action of
the gauged WZW theory (after the gauge field is eliminated)
with the {\it effective} action of the corresponding sigma model.
Then in order to determine the couplings of the sigma model
which appear in its classical action one can ignore all the non-local
terms that are present in the effective action, see Note Added. }

In the next section we shall illustrate the procedure of integrating out  the
gauge field on the example of the $SL(2,R)/U(1)$ gauged WZW model \bn\wit . We
shall find that the target space metric corresponding to the
local part of   $\G (x )$  (or, what turns out to be equivalent,  to $\G' (g)$
restricted  to the gauge-invariant sector) reproduce the exact expression for
the  metric found  in \dvv .  To reproduce the exact expression for the
dilaton one needs to start with the original  non-local effective  action
(3.10),(3.14).

%%%%%%%%%%%%%%%%%%%%%%%%%%%%%
%%%%%%%%%%%%%%
%%%%%%%%%%%%%%%

\newsec {$SL(2,R)/U(1)$ gauged WZW model: field-theoretic derivation of  exact
expressions for  target space fields}
\subsec {Effective action and  exact metric }
 Below we shall demonstrate how the above expressions (3.10),(3.17) for the
effective action  allow one  to obtain the exact form of the corresponding
target space backgrounds in the $SL(2,R)/U(1)$ gauged WZW model.
Let us follow \wit\dvv\ and consider  the
case when  the axial $U(1)$ subgroup of $SL(2,R)$ is gauged.  Using the
parametrisation of $SL(2,R)$ \dvv $$ g= e^{{i\over 2 } \t_L \s_2 } e^{\ha
r\s_1} e^{{i\over 2 } \t_R \s_2  } \ , \ \ \  \ \ \ \ \
  \t \equiv \ha (\t_L - \t_R ) \ , \ \  \tt \equiv \ha ( \t_L + \t_R )\ ,
\eq{4.1} $$ where $r$ and $\t$ are invariant under the axial $U(1)$
transformations (i.e. $x= (r,\t)$)
one can represent the classical action in the form (cf.(3.3))
$$ S(g,A) = k\  [ \ I(g)  - {1\over \pi }
 \int d^2 z \Tr \bigl(  A\,\bd g g\inv +
 \bar A \,g\inv\del g + g\inv A g \bar A  + A \A \bigr)\ ]  \  \eq{4.2} $$
$$ = {k \over 2 \pi } \int d^2 z [ \ \ha  (\del r \bd r  -  \del \t_L \bd \t_L
 -  \del \t_R \bd \t_R - 2 \c \ \bd \t_L \del \t_R )$$
$$  + A(\bd \t_R + \c \ \bd \t_L)  + \A (\del \t_L + \c \
 \del \t_R) - A\A (\c \ + 1) \ ] \ , \eq{4.3}  $$
or, equivalently, in terms of $r, \t , \tt$
$$  S(r, \t , \tt ) = {k \over 2 \pi } \int d^2 z \ \{ \ \ha  \del r \bd r \  +
\ a(r)\  [\ \del \t \bd \t  +  (A- \del \tt ) \bd \t - (\A - \bd\tt ) \del \t
\ ] $$ $$ -\  b(r)  \ (A- \del \tt ) (\A - \bd \tt ) \ \}
\ ,  \eq{4.4} $$
$$ a \equiv \c -1 \ , \ \ \ \ \ b \equiv  \c + 1 \ . \eq{4.5} $$
The gauge field  was assumed to  be proportional to $\ha \s_2$ ( $A$
and $\A$  in (4.2)  are  the   functions  which remain after taking the
trace). Solving for  $A_a$
and substituting it back in the classical action one finds that the gauge
degree
of freedom $\tt$  drops out  and  so that one  is left with the $D=2$ model
 \wit\bcr\
$$ S(r, \t) =  {k \over 4 \pi } \int d^2 z \ [\ \del r \bd r  + 4
\th^2 {r\over 2} \
 \del \t \bd \t \ ]  \ . \eq{4.6} $$
The effective action (3.10),(3.17)   for  the case  when one
gauges an abelian   vector subgroup  has
 an   obvious analog  in   the case of the gauging of an abelian axial
subgroup.   The local part of the effective action (3.17)  (corresponding to
the gauge $\del^a A_a =0$)  takes the form (in the present model $c_G=-4, \ c_H
=0$)
 $$\Gl(g,A) = (k-2) \big[ \ I (g, A )
-\1p { 4 \over  (k-2) } \in \Tr (   A\A  ) \ \big]
 $$
$$ = (k-2) \big[ \  I_0 (g,A) - \1p (1 +  { 4\over k-2 } )
\in \Tr (   A\A  ) \ \big] \ \ . \eq{4.7}  $$
Explicitly (cf.(4.3),(4.4))
$$ \Gl(g,A)=
 {(k-2) \over 2 \pi }\int d^2 z\  [ \ \ha  (\del r \bd r  -  \del
\t_L \bd \t_L
 -  \del \t_R \bd \t_R  - 2 \c \ \bd \t_L \del \t_R)$$
$$  + A(\bd \t_R + \c \ \bd \t_L)  + \A (\del \t_L + \c \
\del \t_R) - A\A (\c \  + 1 + {4\over k-2 } ) \ ] \  \eq{4.8} $$
$$  ={(k-2) \over 2 \pi }\int d^2 z \ \{ \ \ha  \del r \bd r \  + \ a(r)\
[\ \del \t \bd \t  +  (A- \del \tt ) \bd \t - (\A - \bd\tt ) \del \t \ ]
$$ $$ - \ b(r)  (A- \del \tt ) (\A - \bd \tt ) - {4\over k-2 } A\A \ \}
\ .  \eq{4.9} $$
Since  the local part of the effective action is no longer
invariant under the (non-holomorphic)  axial transformations (because of the
``quantum correction" ${4\over k-2 }$  in the coefficient  of the $A\A$ term)
the field $\tt$ will no longer automatically decouple upon elimination of $A$
from the action.

If we restrict consideration to ``gauge invariant sector"
$\tt=0$   then integrating  out $A$ we  will get the following exact
generalisation of the $D=2$  semiclassical euclidean black hole action
(4.6)\foot { If one
 integrates out $A$ starting directly from the non-local effective action
(3.10) one obtains an action which  contains non-local terms in $r$ but
coincides with (4.10) for $r=\const$ (see below).}
$$ \G (r, \t) =
{(k-2 )\over 4 \pi } \int d^2 z  \ [ \ \del r \bd r  +  f (r)
 \del \t \bd \t \ ]  \ , \ \ \ f(r)={ 4\th^2 {r\over 2} \over
 1-{2\over k} \th^2 {r\over 2}}\ .  \eq{4.10} $$
In this way we reproduce the exact metric of ref.\dvv .\foot {Note that
   \dvv\  also used the restriction equivalent to $\tt =0$ in  computing
the $L_0$-operator. }

In general the  action (4.9) can be rewritten in the form
$$  \Gl = {(k -2)\over 2 \pi } \int d^2 z \ \{ \  \ha  \del r \bd r  + a \del
\t \bd \t  + a ( B \bd \t -  \B  \del \t )  -  b B\B - \g  (B+ \del
\tt ) (\B + \bd \tt ) \ \}  \ , \eq{4.11} $$
 $$  \ \ \ \ B\equiv A - \del \tt \  , \ \ \ \B\equiv \A - \bd \tt \ , \ \ \
\ \ \ \g = {4\ov k-2 } \  .   $$ Eliminating $B, \B$ from the action we get
$$ \G (r , \t , \tt ) = {(k -2)\over 4 \pi } \int d^2 z
\ [  \   \del r \bd r  +  f (r)  \del \t
\bd \t  +  E(r) (  \bd \tt \del  \t -  \bd \t \del \tt )  -  H (r)\del \tt \bd
\tt \ ]\ , \ \eq{4.12} $$
$$ f = 2( a -  { a^2 \ov b + \g }) =  { 4\th^2 {r\over 2} \over
 1-{2\over k} \th^2 {r\over 2}}\ \ , \eq{4.13} $$
$$ E = { 2\g a \ov b + \g } = {2\ov k } f \ \ , \ \ \ \ \
H={ 2 \g b \ov b + \g } =  {4\ov k^2 } f + {8 \ov  k }=
{8 \ov  k ( 1-{2\over k} \th^2 {r\over 2}) }\ .
\eq{4.14} $$ The action (4.12) generalizes (4.10) to the case of $\tt \not= 0
$  ($\tt$ disappears from  (4.12)  in the limit $k \ra \infty$ in which the
action reduces to (4.6)).
It is interesting to note that  (4.12)  coincides with the
``charged black hole" action or semiclassical action of the $(SL(2,R)\times
U(1)) / U(1) $ model \ishi \rocver\ (which is  related by a duality
transformation to  the semiclassical action of the ``black string" model
$(SL(2,R)/U(1)) \times U(1)$  with  a  particular value of the ratio of the
mass to the charge  corresponding to the value  $ \l={- 2\ov k-2}$ of the
parameter  in \rocver, see (A.6)).  The action (4.12) can be represented also
in the following  form $$ \G (r , \t , \tt ) = {(k -2)\over 4 \pi } \int d^2
z  \ [  \   \del r \bd r  +  G   \del (\t-  \tt' )\bd (\t +  \tt' )
 - 2k  \del \tt' \bd \tt' \ ]\  \
,  \eq{4.15} $$ $$   \ \ \tt' \equiv {2\over k } \tt \ \ ,    $$
 or, equivalently,
      $$ \G (r , \t , \tt ) = {(k -2)\over 4 \pi } \int d^2 z
\ [  \   \del r \bd r   - 2k \del \t \bd \t + {k^2 \ov 4 } H
   \del (\t-
\tt' )\bd (\t +  \tt' ) \ ] \ . \eq{4.16} $$
We see that for $r \not= \const$  the field $ \tt$ does not decouple from
$\t$ and hence cannot be easily integrated out: integration
over $\tt$ gives extra non-local $O(\del r )$ - dependent terms.

Let us now consider  the integration over $A_a$ in the case
when one starts directly  with the   original non-local  effective action
(3.10),(3.14), i.e.  does not   first impose  the  gauge condition on $A_a$
which reduces
 $\G (g,A) $ to $\Gl (g,A)$.
Introducing the fields $\r_L$ and $\r_R$ (corresponding to $h$ and $\bh$ in
(3.4))
 $$ A=\del \r_L\ \ , \ \ \ \A=\bd \r_R\ \ , \ \ \  \r = \ha (\r_L-\r_R)\ \
,  \ \ \ \tr = \ha (\r_L + \r_R)\ \ , \ \eq{4.17} $$
where  $\r$ and $\tr$ represent the transverse and  longitudinal parts of
$A_a$,   we  get for    the classical action (4.4)
  $$ S(r, \t ,\tt ,\r , \tr )= {k \over 2 \pi }
 \int d^2 z \ \{ \ [ \ \ha  \del r
\bd r  -  a \del \k \bd \k
 -  b  \del \tk \bd \tk   +  a   (\del \k \bd \tk - \bd \k \del \tk) \ ] \
+\  2 \del \r \bd \r \ \} \ , \eq{4.18} $$
$$ \k \equiv \t + \r \ \ , \ \ \ \ \tk   \equiv \tt \ - \tr . $$
The gauge invariance manifests itself in   that the action depends  on only one
combination $\tk$ of  the longitudinal part  of $A_a$  ($\tr$) and  $\tt$.
The two groups of terms in (4.18) correspond to the two terms in (3.5) (the
terms in the square bracket  represent the action of the $SL(2,R)$ WZW
model).
The effective action  (3.10),(3.14) then takes the form
$$ \G (r, \t ,\tt ,\r , \tr )= {(k-2) \over 2 \pi } \int d^2 z\  [ \ \ha  \del
r
\bd r  -  a\del \k \bd \k
 -  b\del \tk \bd \tk $$ $$ +   a   (\del \k \bd \tk - \bd \k \del \tk)
+ {2k\over k-2}   \del \r \bd \r \ ] $$
$$ = {(k-2) \over 2 \pi } \int d^2 z \ [ \ \ha  \del r \bd
r  -  a\del \k \bd \k
 -  b \del \tk \bd \tk   $$ $$  + a   (\del \k \bd \tk - \bd \k \del \tk)
+ {2k\over k-2}  (  \del \k \bd \k - 2 \del k \bd \t + \del \t \bd \t )  \ ] \
... \eq{4.19} $$
 Note that   it is the
dependence on the transverse part $\r$ of $A_a$ that is  modified  in the
effective action.   Eliminating  $\k$ from the action we obtain (4.12) where
$\tt$  should be identified with $\tk$ (recall that in deriving (4.12) we have
used the gauge  $\tr$=0).  Integrating over $\k$ and $\tk$  we get  (4.10) up
to non-local terms depending on derivatives of $r$.

  It is possible to represent the result of elimination of $A_a$ in terms of  a
local action for the
 $three$ fields: $r,\t$ and an  extra field $\hi$, i.e. in the form
 similar to  (4.15),(4.16).   Introducing  an
auxiliary  scalar field $\hi$   one can replace  the effective action (3.14)
$$ \G (r, \t , A ) =   {(k-2)
\over 2 \pi } \int d^2 z\  [ \ \ha  \del r \bd r  -  b(r) A\A  $$ $$  -
{1\over k-2 } F { 1\over \del \bd } F + {2k\over k-2} F \t   +
 {2k\over k-2}   \del \t \bd \t   \ ] \eq{4.20} $$
by  the  equivalent local functional
$$  \G (r, \t , A, \hi )= {(k-2) \over 2 \pi
} \int d^2 z \ [ \ \ha  \del r \bd r  -  b(r) A\A   $$ $$  -  {1\over k-2 }
\del \hi \bd \hi  + {2\over k-2} F ( k\t +   \hi )   +
 {2k\over k-2}   \del \t \bd \t   \ ] \ .  \eq{4.21} $$
We have fixed the gauge $\tt=0$ (in any case  $\tt$   drops out
after the integration over $A_a$ because of the gauge invariance  of $\G$).
Now the  elimination of  $A,\A$ from the action becomes  straightforward and
we are left with  $$ \G (r, \t , \hi ) = {(k-2) \over 2 \pi } \int d^2 z \ [ \
\ha  \del r \bd r   + {2k \over k-2} \del \t \bd \t  -  {1\over k-2 } \del \hi
\bd \hi  $$ $$ - {4\over (k-2)^2 b(r) } \del ( \hi + k \t ) \bd (\hi + k \t )
\ ] \ . \eq{4.22}  $$
Integration over $\hi$ leads us back to  (4.10) up to non-local  $O(\del r )$
terms.

We conclude that  the local part of the effective  $D=2$ action
for $r$ and $\t$  is given by (4.10).\foot { The effective action is manifestly
local only in terms of $D=3$ representation (4.15),(4.16) or (4.21).}
We thus reproduced the exact  background of ref. [9] in the field-theoretic
approach.
The  local $D=2$ model (4.10)   (with  the dilaton
coupling to be discussed below) is   Weyl invariant   (as was checked
 explicitly up to four
leading  orders in loop expansion  \ts\Jack).

%%%%%%%%%%%%%%%%%%%%%%%%%%%%%%%%%%%%%%%%%
%%%%%%%%%%%%%%%%%%%%%%%%%%%%%%%%%%%%%%%%%
%%%%%%%%%%%%%%%%%%%%%%%%%%%%%%%%%%%%%%%%%%%%%%%%%%%%%%%%%%%%%%%%%%%%%
\subsec { Dilaton  field and measure factor}
Though the ``naive"  form   of  the effective action ((4.9) with $\tt
=0$) is sufficient in order to obtain the  expression for the metric
(4.10),   to derive the  exact  form  of the dilaton  coupling    it is
necessary to start with the full gauge invariant  non-local effective action
(3.10),(3.14) or (4.19),(4.20) defined on a curved  $2d$ background. This
provides an important check of the consistency of  our approach.

As in the leading order approximation {\wit } the dilaton term originates from
the determinant resulting from  the integration over  the gauge field.
This determinant  is  not   unambiguously defined (see \shts ).  The freedom of
adding local counterterms  should be fixed by using the condition  that
conformal invariance of the original theory should be preserved in the process
of integrating out a subset of fields.\foot {Put differently, conformal
invariance of the
original theory holds  (without  need to make additional field redefinitions)
only within a  specific  regularisation  scheme; it is that  particular
regularisation  that should be used in computing the determinants resulting
from
integration over some fields.}  The correct definition of the determinant
(both at the semiclassical and exact levels) corresponds to treating $A_a$ as
being built out of the scalar fields $\r,\tr$ (4.17) with the latter
considered as   fundamental integration variables \bush\mplt\shts \foot {Note
that the effective action is local being expressed (4.19) in terms of $\r$
and $\tr$.}.

  Let us first discuss the case which
corresponds to integrating out the vector field in the classical action  (4.3).
 Consider  the following  integral over the $2d$ vector field $A_a$
$$ Z= \int [dA_a] \exp {[ - \2p \int d^2z \sqrt g \ M(z) A_a A^a  }
\ ]\ \ ,
 \eq{4.23} $$
where $M$ is a given  function.  This integral is not well defined \shts .
 Using
different definitions (regularisations) of (4.23) one will get   expressions
which will  differ by local counterterms,
$$  Z = \exp [ -  ({\rm Tr } \ln M )_{reg} ]  $$ $$ =
\exp [ -{1\over { 8 \pi  }} \int
d^2 z \sqrt {g} \ [
 c_0 \Lambda^2  \ln M  +  c_1
(\del_a \ln M )^2  \  +  \  c_2  R \ln M  \ ] \ \ , \  \eq{4.24}
$$  where $\Lambda $ is an UV cutoff, $R$ is the  curvature  of $2d$ metric and
$c_i$ are finite coefficients.
 A choice of a particular definition of $Z$ is dictated by some additional
conditions which the total theory  should
satisfy.  In the present case the  preservation of the conformal invariance
demands that  $c_1=0$ and $c_2= -1$ \bush\mplt\shts .
The quadratically divergent  term in (4.24) can be identified  with
a contribution to  a  local measure  (regularized in a naive way)  so that
(4.24) can be represented in the form
$$ Z= \prod_z M^{-1}(z) \ \exp \  [  \ {1\over {
8 \pi  }} \int d^2 z\  \sqrt {g}  \ R \ \ln M (z)  \ ] \ \ . \ \ \eq{4.25} $$
It is important to stress that  (4.25) is the product of two $separate$
factors:  a local measure factor and  a (global) dilaton contribution, $$ \p =
- \  \ha  \ln M  \ \ . \eq{4.26} $$
Note that  since
$$ \e{2\p (z) } =  M^{-1} (z) \ \  \eq{4.27} $$
one may  mis-interpret  the measure factor  as being related to  the
dilaton  contribution (or vice versa). The two factors play, in fact, very
different roles and should not be mixed up.
   For example, the dilaton
factor is absent on a flat $2d$ background while  the measure factor  must be
present irrespective of the   value of the curvature of $2d$ metric (the
measure
factor is important for cancellation of quadratic divergences  in the  total
theory while the dilaton contribution is essential for  the absence  of the
Weyl anomalies).\foot {
 There  seems to  exist   some confusion
in the literature concerning   a relation between  the dilaton  term  and the
measure factor. The above remarks hopefully clarify the
issue. }

 In the case of the  $SL(2,R)/U(1)$ model treated in the
semiclassical approximation it  is the first  factor in (4.25)  that combines
with the Haar measure on $SL(2,R)$ to give the $\sqrt G$ measure factor of the
resulting $D=2$ sigma model (4.6). In fact,  after fixing the gauge $\tt =0$
and integrating out the gauge field
the respective measure factors are  given by
$$  d\m (r,\t ) =  \prod_z dr(z) \ d\t (z) \ \sh r(z)  \ \prod_z  M^{-1}(z)
$$ $$ = N  \prod_z dr(z) \ d\t (z) {\sqrt{ G(z)}} \ , \eq{4.28} $$ $$    M=
k(\c + 1 ) \ , \ \ G = 4k^2 \th^2 { r\over 2} \ , \ \ N =\ha  k^{-2}\ ,  $$
so that the resulting sigma model partition function is ($x^\mu = (r,\t)$)
$$ Z_{s.m.} = N \int \prod_z dx^\mu (z){\sqrt { G(z)}} \exp [
- {1\over { 4 \pi  }} \int d^2 z
\sqrt {g} \ (G_{\m \n} \del_a x^\m
\del^a x^\n \  +  \  R  \p  ) \ ] \  . \eq{4.29}  $$
Let us now see how the same analysis
goes through if we start with the  effective action (4.19),(4.20). Since the
squares of longitudinal and  transverse parts of $A_a$ have different
coefficients in the effective action, i.e. $$ M A^2_a \ra  M_{\Vert }
A_{\Vert }^2 + M_\bot A_\bot^2 \ \ , $$ $$
 M_{\Vert } = (k-2) (\ \c + 1) \ , \ \ \ \ \  M_\bot = (k-2) (\ \c + 1 +
{4\over k-2} ) \ , \eq{4.30} $$
the result of integration  over  $A_a$ (i.e. over its transverse and
longitudinal parts $\r$ and $\tr$) is given by  (4.25) with $M$ replaced by
$\sqrt {M_{\Vert } M_\bot }$, i.e.\
 $$ Z= \prod_z M_{\Vert }^{-\ha}(z) M_\bot^{-\ha}(z) \ \exp \ [
{1\over { 16 \pi  }} \int d^2 z \ \sqrt {g} \  R \ \ln ( M_{\Vert } (z)  M_\bot
(z) )\  ] \ .  \eq{4.31} $$
Then  the exact form of the measure factor (eq.(4.28)) is
$$  d\m (r,\t ) =  \prod_z dr(z)\  d\t (z) \ \sh r(z) \
 \prod_z M_{\Vert }^{-\ha}(z) M_\bot^{-\ha}(z) $$ $$
= N  \prod_z dr(z) \ d\t (z) {\sqrt {G(z)}} \ ,  \eq{4.32} $$ $$ \  \ \ G =
(k-2)^2 { 4\th^2 {r\over 2} \over
 1-{2\over k} \th^2 {r\over 2}}\ , \ \ \ \ N =\ha k^{-1/2}(k-2)^{-3/2}
\ .  $$
The dilaton contribution is given  by
 $$ \p =\p_0 - \fourth \ln  M_{\Vert }  -
\fourth \ln  M_\bot  = \p_0{ }' -  \ha \ln  \sh r
+ \fourth \ln  G \ , \eq{4.33}  $$
i.e we reproduce the expression of \dvv .

Note that  since
$$ \e{2\p  } =  M_{\Vert }^{-\ha}  M_\bot^{-\ha} \ \eq{4.34} $$
it follows from (4.32)  that
$$  \ \  \sqrt G \ \e{-2 \p} = N^{-1} \sh r \ .
\eq{4.35} $$
This implies that in general  $\sqrt G \ {\rm e}^{-2 \p}$  being proportional
to the Haar measure (with proper gauge fixing) should be essentially
$k$-independent (up to an overall  constant factor) in agreement with previous
suggestions \kir\bsft.\foot { I am grateful to I. Bars and K. Sfetsos  for
an important discussion of the $k$-independence of the factor (4.35). }  It  is
not clear if there is more than  just a coincidence in the fact that  $\sqrt G
\ {\rm e}^{-2 \p}$  is, at the same time, the  zero mode measure factor which
appears in the sigma model partition function defined on  a 2-{\it sphere } or
in the corresponding  tree level string effective action \frts  $$ Z\sim \int
d^Dx \ \sqrt G\ {\rm e}^{-2 \p} \ .  $$

 \newsec {Concluding remarks}
As we have seen in Sec.4 the local part of the
  effective action of the $SL(2,R)/U(1)$ model
is given by the sigma model action (4.10)
with the exact couplings of ref. [9]. The full effective action
is local if  expressed in terms of  gauge invariant components $(r, \t)$ of
$g$
 $and$ an extra scalar field.  The latter  is also gauge invariant and can be
identified  either   with  $\tk$ in (4.18) (or, equivalently, $\tt$ in (4.12))
or with an axiliary field $\hi$ in (4.21). This  field  disappears from
the action in the limit $k \ra \infty$ , i.e. it is absent in the leading order
approximation.  Once we  integrate over this extra field (i.e. integrate over
all components  of the gauge field) we  get the $D=2$     action (4.10)
for $x=(r,\t)$  plus   non-local  $O(\del r)$ terms
(which vanish in the $k \ra \infty $ limit ).

The approach suggested in this paper
 should find applications in  similar  models.    In particular, it is
important  to  apply analogous  method  to  derivation of  exact backgrounds
\bsfet\
 corresponding to  $G/H$ gauged WZW models  with {\it non-abelian} subgroups
$H$ \bs\ttt.

It is  straightforward to  generalise  the  above   analysis     to
the    supersymmetric  gauged WZW model \ttt.
  In the supersymmetric  case there is
no shift of $k$  in  the  effective action  of WZW model (2.11) (because of an
extra contribution of fermions \div). The effective action
of the supersymmetric gauged WZW model is thus equal to the classical one
(cf.(3.5),(3.9)) so that there are no $1/k$ corrections to the leading
order form of the background fields in agreement with  what one finds using
the $L_0$ - argument \Jack\bsfet .

\bigskip \bigskip
I am very grateful to   I. Bars and K. Sfetsos for  stimulating
and useful discussions and collaboration. I would like  to thank   C.
Hull, V. Fateev and  M. Shifman  for   helpful remarks.  Part of this work
was done at Aspen Centre for Physics. I   would like also to acknowledge the
support of SERC. \bigskip

\bigskip\bigskip
\centerline{ Note Added}
The approach of this paper was further  developed and clarified
in our recent work \ttt.   In particular, we   pointed out that
one should identify the effective action of the gauged WZW  theory
with the effective action of the corresponding sigma model. Then to find
the sigma model couplings one needs  to consider only the local part of the
effective action. As a result, one can ignore non-local terms originating from
the field renormalisations in (2.11) and (3.9). It is possible also to
truncate the  quantum term (3.11) in the effective action (3.10) to
its quadratic part (3.15). As a result,  one is able to derive
the general expressions for the sigma model couplings (metric, dilaton and
 antisymmetric tensor) in the case of an
 arbitrary non-abelian subgroup $H$ (see also \bs).

\bigskip\bigskip
%%%%%%%%%%%%%%%%%%%%%%%%%%%%%%%%%%%%%%%%%%%%%%
\newsec {Appendix }
Below we shall consider   several actions related to the actions from Sec.4
by duality transformations.  In particular, we shall show that the
effective action of the $SL(2,R)/U(1)$ model
(4.12) is related through duality transformations to the action of the ungauged
$SL(2,R)$ WZW model.  Since the leading order form of the duality
transformation which we shall use  preserves  conformal invariance only to one
loop order the resulting actions  are conformal invariant only  in the one
loop approximation and should not be expected to represent exact backgrounds.
Even though making a duality transformation may simplify the structure of the
actions  (4.12),(4.22) the property of  conformal invariance to all loop
orders, in general,  may  be lost.

 Let us consider the  sigma model
$$I= {1\over { 4 \pi  }} \int d^2 z \sqrt {g} \ [ (G_{\m \n} + B_{\m \n}) (
g^{ab} + i \ep^{ab}) \del_a x^\m \del_b x^\n \  +  \  R  \p \ ] \  \ ,
\eq{A.1} $$
which is invariant under an abelian  isometry $\d x^\m = \ep K^\m $.
Choosing  coordinates  $ \{ x^\m \} = \{ x^1 \equiv y \ , \  x^i \} $
in such a way that $G_{\m \n}$, $B_{\m \n}$  and $ \p $ are independent of  the
coordinate $y$ which is shifted by the isometry   one finds  the
dual action $\tilde {I}$
for  $ \{ \tx^\m \} = \{ \tx^1 \equiv \ty \ , \  x^i \} $
 which has the form (A.1)  with  \bush
$$ {\tilde G}_{11} = M^{-1} \ \ , \ \ {\tilde G}_{1i} =  M^{-1} B_{1i} \ \
, \ \ {\tilde G}_{ij} =  G_{ij} - M^{-1} ( G_{1i} G_{1j}- B_{1i} B_{1j})\ \ ,
$$ $$
\ \ {\tilde B}_{1i} =  M^{-1} G_{1i} \ \
, \ \ {\tilde B}_{ij} =  B_{ij} - M^{-1} ( B_{1i} G_{1j}- G_{1i} B_{1j})\ \ ,
\ \ \ M \equiv G_{11} \ \ , \ \eq{A.2} $$
$$ {\tilde \phi } = \phi - \ha  \ln M  \ \ . \eq{A.3} $$
Applying this transformation  to the action of the ungauged
$SL(2,R)$ WZW model (i.e. (4.4),(4.18) for  $A=0$)  with $y = \tt$
$$ S(r, \t ,\tt  )= {k \over 2 \pi }
 \int d^2 z  \ [ \ \ha  \del r
\bd r  -  a(r) \del \t \bd \t
 -  b(r)  \del \tt \bd \tt   +  a (r)  (\del \t \bd \tt - \bd \t \del \tt) \ ]
 \ , \ \ \  \p = \p_0 \ , \eq{A.4} $$
we find \kumar\
$$  {\tilde S} (r, \la , \s ) =   {k \over 2 \pi } \int d^2 z \  [ \ \ha  \del
r \bd  r  + 2 \th^2 {r\over 2}  \del \la \bd \la -  2\del \s \bd \s  ] \ , \ \
\p = \p_0{}' -  \ha \ln ( \ch { r}  + 1 ) \ , \eq{A.5} $$
 where $\la$ is the field dual to  $\tt$
and $\s\equiv  \t - \la $.  This model corresponds to the ``black string"
background \hor, i.e. is the direct product of the  leading order
$D=2$ black hole (4.6) and an extra free field.  Though the $SL(2,R)$  WZW
model (A.4) we started with is conformal to all orders in $\a'$ expansion the
resulting model (A.5) is conformal only to the leading order in $\a'$.  As we
have mentioned above, this is not surprising since  (A.2),(A.3) represent only
the leading order form of the duality transformation \mplt.

 The action (A.5) has two commuting isometries. Acting   on  (A.5) by  a
particular $O(2,2)$ transformation \rocver\  we get  the
(euclidean) charged black hole metric \ishi\foot { The conformal field
theories corresponding to the charged black hole and  a compact black
string are thus equivalent \rocver. }
$$  S=   {k \over 2 \pi } \int d^2 z\  [ \ \ha  \del r \bd  r  + {(1-\l)\sh^2
{r\over 2} \over \ch^2 {r\over 2} - \l }  \del \la_1 \bd \la_1
$$ $$ +{\l \ \sh^2
{r\over 2} \over k (\ch^2 {r\over 2} - \l )  }( \del \la_1 \bd \la_2 -
\del \la_2 \bd \la_1 )    + {\l \ \ch^2
{r\over 2} \over k^2 (\ch^2 {r\over 2} - \l  )  }  \del \la_2 \bd \la_2 \  ]
\ \ ,  \eq{A.6} $$
$$ \p = \p_0{}''  -  \ha \ln ( \ch { r}  + 1  - 2 \l ) \ , \eq{A.7} $$
Here $\l $ is  a free parameter related to the charge of black hole.
 If we now take $ \l = - {2\over k-2} $ (A.6) becomes equivalent to the
effective action (4.12). The dilaton factors are also equivalent as one
can see by noting that the dilaton field corresponding to (4.14)  originates
from the integral over $B,\bar B$ and thus is given by $
 \p = \p_0  -  \ha \ln [ b(r) + \g  ]$ (see (4.26)).
 As a result,
we discover that  the $\la_2 =0$ section of the
 metric  corresponding to (A.6) coincides with  the exact $D=2$  black hole
metric   (4.10) ($ {(1-\l){\rm sinh}^2 {r\over 2} \over {\rm cosh}^2 {r\over 2}
- \l } = {  \th^2 {r\over 2 } \over 1 - {2\over k } \th^2 {r\over 2 } }
)$.\foot { It is interesting to note that though we have used the leading order
form of the duality transformation we got the exact $D=2$ black hole  metric
in the $\la_2 =0$ section.}

Note that the action (4.22) does not contain the  antisymmetric tensor  term.
It can be generated by making a duality transformation of $\hi$.  One should be
careful, however, not to mix  $\t$ and $\hi$ by field redefinitions since  it
is $\hi$ that should be eventually integrated out  to obtain a $D=2$ model.

\listrefs
%\listfigs   %(if necessary)
\bye